\documentclass[epj]{svjour}
\usepackage{graphics}
\begin{document}
\title{Problems with Fitting to the Power-Law Distribution}
\author{Michel L. Goldstein\inst{1} \and Steven A. Morris\inst{2} \and Gary G. Yen\inst{3}
}                     
\institute{School of Electrical and Computer Engineering, Oklahoma State University, Stillwater, OK 74078}
\date{Received: date / Revised version: date}
%
\abstract{ This short communication uses a simple experiment to show
that fitting to a power law distribution by using graphical methods
based on linear fit on the log-log scale is biased and inaccurate.
It shows that using maximum likelihood estimation (MLE) is far more
robust. Finally, it presents a new table for performing the
Kolmogorov-Smirnof test for goodness-of-fit tailored to power-law
distributions in which the power-law exponent is estimated using
MLE. The techniques presented here will advance the application of
complex network theory by allowing reliable estimation of power-law
models from data and further allowing quantitative assessment of
goodness-of-fit of proposed power-law models to empirical data.
} 
\maketitle
\section{Introduction}
\label{intro} In recent years, a significant amount of research has
focused on showing that many physical and social phenomena follow a
power-law distribution. Some examples of these phenomena are the
World Wide Web \cite{key-25}, metabolic networks \cite{key-29},
Internet router connections \cite{key-27}, journal paper reference
networks \cite{key-40}, and sexual contact networks \cite{key-35}.
Often, simple graphical methods are used for fitting the empirical
data to a power-law distribution. Such graphical analysis, based on
linear fitting of log-log transformed data, can be grossly
erroneous.

The pure power-law distribution, known as the zeta distribution, or
discrete Pareto distribution \cite{key-30} is expressed as
\begin{equation}p(k)=\frac{k^{-\gamma}}{\zeta(\gamma)},\end{equation}
where:
\begin{itemize}
\item $k$ is a positive integer usually measuring some variable of interest, e.g.,
number of links per network node;
\item $p(k)$ is the probability of observing the value $k$;
\item $\gamma$ is the power-law exponent;
\item $\zeta(\gamma)$ is the Riemann zeta function defined as $\displaystyle \sum_{k=1}^{\infty}k^{-\gamma}$.
It is important to note, from this definition, that $\gamma>1$ for
the Riemann zeta function to be finite.
\end{itemize}

Without a quantitative measure of goodness-of-fit, it is difficult
to assess how well data approximates a power-law distribution.
Moreover, a quantitative analysis of the goodness-of-fit enables the
identification of possible interesting phenomena that could be
causing the distribution to deviate from a power-law. In some cases
the underlying process may not actually generate power-law
distributed data, which may instead be due to outside influences,
such as biased data collection techniques or random bipartite
structures \cite{key-28}. Quantitative assessment of the
goodness-of-fit for the power-law distribution can assist in
identifying these cases.

This paper demonstrates that the current broadly used methods for
fitting to the power-law distribution tend to provide biased
estimates for the power-law exponent, while the maximum likelihood
estimator (MLE) produces more accurate and robust estimates.
Finally, MLE permits the use of a Kolmogorov-Smirnov (KS) test to
assess goodness-of-fit. This paper provides a new KS table suitable
for testing power-law distributions derived from MLE estimation.

\section{Problems of currently used estimation methods}
\label{sec:1} In the literature, many researchers make parameter
estimations using simple graphical methods, such as 1) direct linear
fit of the log-log plot of the full raw histogram of the data
\cite{key-1,key-36}, 2) fit of the first 5 points of the log-log
plot of the raw histogram \cite{key-31}, or 3) linear fitting to
logarithmically binned histograms \cite{key-25,key-39}. The easy
graphical nature of these methods tends to mask their basic
inaccuracy. In a simple experiment, a random deviate generator was
used to produce a dataset of 10,000 samples from a known zeta
distribution with exponent $\gamma=2.500$. The three graphical
methods listed above were used to estimate the power-law exponent
from the dataset. This experiment was repeated 50 times and the
tabulated results are presented in Table~\ref{tab:1}. Linear fitting
was performed using least squares regression, where the slope of the
fit was used as the estimate of the exponent $\gamma$. MLE estimates
of the exponent are also included in the table.

\begin{table}
\caption{Sample results of parameter estimation using various
methods for 10,000 samples of power-law distribution with $\gamma =
2.500$. Sample result based on 50 runs.}
\label{tab:1}       
\begin{tabular}{lccc}
\hline\noalign{\smallskip}
 & Mean & & \\
 & estimated & & Bias \\
Estimation method & $\gamma$ & $\sigma$ & error  \\
\noalign{\smallskip}\hline\noalign{\smallskip}
Linear & 1.590 & 0.184 & 36\% \\
Linear 5-points & 2.500 & 0.045 & 0 \\
Log-2 bins & 1.777 & 0.038 & 29\% \\
MLE & 2.500 & 0.017 & 0 \\
\noalign{\smallskip}\hline
\end{tabular}
\end{table}

This table shows that two of the methods, full linear fit, and
linear fit on logarithmic bins, suffer from severe bias, with 36\%
and 29\% bias error respectively. The most accurate of the three
graphical methods is the linear fit of the first 5 points, where the
estimate is based on the slope of the first 5 points of the
distribution histogram in log-log scale. These first 5 points
contain most of the data and, due to the large number of samples,
they can decrease the bias caused by the large uneven variation in
the tail (the log-log transformation distorts the error in the tail
unevenly). However, the variance of this estimate is much higher
than the variance of estimates from MLE, showing the stability of
MLE.

Maximum likelihood estimation of the zeta distribution \cite{key-30}
maximizes the log-likelihood function given by, assuming
independence between the data points:

\begin{eqnarray}
l(\gamma\mid x) & = &
\prod_{i=1}^{N}\frac{x_{i}^{-\gamma}}{\zeta(\gamma)}\nonumber
\\
L(\gamma\mid x) & = & \log l(\gamma\mid x)\nonumber
\\
&
 = & \sum_{i=1}^{N}(-\gamma\log(x_{i})-\log(\zeta(\gamma))) \nonumber\\
 & = & -\gamma\sum_{i=1}^{N}\log(x_{i})-N\log(\zeta(\gamma)),
\end{eqnarray}
where:

\begin{itemize}
\item $l(\gamma\mid x)$ is the likelihood function of $\gamma$ given the
unbinned data $x={x_i}_{1<=i<=N}$,
\item $L(\gamma\mid x)$ is the log-likelihood function.
\end{itemize}
This maximum can be obtained theoretically for the zeta distribution
by finding the zero of the derivative of the log-likelihood function

\begin{eqnarray}
\frac{d}{d\gamma}L(\gamma\mid x) & =
& -\sum_{i=1}^{N}\log(x_{i})-N\frac{1}{\zeta(\gamma)}\frac{d}{d\gamma}\zeta(\gamma)=0 \nonumber\\
\Rightarrow\frac{\zeta\prime(\gamma)}{\zeta(\gamma)} & = &
\frac{1}{N}\sum_{i=1}^{N}\log(x_{i}) \label{eq:zz}
\end{eqnarray}

where: $\zeta\prime(\gamma)$ is the derivative of the Riemann Zeta
function.

A table with the value of the ratio
$\zeta\prime(\gamma)/\zeta(\gamma)$ can be found in \cite{key-41},
or values can be generated on most modern mathematical and
engineering calculation programs such as Matlab, Maple and
Mathematica.

Note that the parameter estimate of a power-law exponent has very
limited meaning without some assessment of its goodness-of-fit. The
KS test is a robust and simple goodness-of-fit test that can be used
to obtain this information.

\section{Using a KS-Type Goodness-of-Fit Test for Power-Law Distribution Hypothesis}
\label{sec:2} The two most commonly used goodness-of-fit tests are
Pearson's $\chi^{2}$ test, and the Kolmogorov-Smirnov (KS) type
test. The Pearson's $\chi^{2}$ test is very simple to perform but
has severe problems related to the choice of the number of classes
to use \cite{key-37}. Because of this, in most cases it is
preferable to use the KS test. The KS test is based on the following
test statistic:

\begin{equation} K=\sup_{x}\left|F^{\star}(x)-S(x)\right|,\end{equation}
where:
\begin{itemize}
\item $F^{\star}(x)$ is the hypothesized cumulative distribution function
\item $S(x)$ is the empirical distribution function based on the
sampled data.
\end{itemize}

Kolmogorov \cite{key-32} first supplied a table for this test
statistic for the case where the hypothesized distribution function
was independent to the data, i.e., when none of the parameters of
the hypothesized distribution function is extracted from the data.
When there is a dependency, other tables must be used. This
limitation was not taken into consideration by Pao and Nicholls in
their application \cite{key-37,key-38} of the KS test to power-laws.
Without correcting for this factor, the KS test gives a rejection
rate lower than what is expected \cite{key-26}.

Lilifoers later introduced tables for using the KS test with other
distributions, such as normal and exponential \cite{key-33,key-34}.
These tables were obtained using a Monte Carlo method, which is
based on generating a large number of distributions with random
parameters and calculating the test statistic for each of the test
cases, from which empirical values for the quantiles can be
extracted. The same procedure was used to obtain these values for
the power-law distribution. For each of logarithmically spaced
sample sizes, 10,000 power-law distributions were simulated, with
random exponent from 1.5 to 4.0. Statistics were collected from
these simulations to generate the KS quantiles, shown in Table
~\ref{tab:2}. This table was created assuming MLE as the estimation
method. Separate KS tables would have to be constructed for other
estimation methods.
\begin{table}
\caption{KS test table for power-law distributions, assuming MLE
estimation.}
\label{tab:2}       
\begin{tabular}{lllll}
\hline\noalign{\smallskip} &
\multicolumn{4}{c}{Quantile} \\
\# samples & 0.9 & 0.95 & 0.99 &
0.999  \\
\noalign{\smallskip}\hline\noalign{\smallskip} 10 & 0.1765 & 0.2103
&
0.2835 & 0.3874 \\
20 & 0.1257 & 0.1486 & 0.2003 & 0.2696 \\
30 & 0.1048 & 0.1239 & 0.1627 & 0.2127 \\
40 & 0.0920 & 0.1075 & 0.1439 & 0.1857 \\
50 & 0.0826 & 0.0979 & 0.1281 & 0.1719 \\
100 & 0.0580 & 0.0692 & 0.0922 & 0.1164 \\
500 & 0.0258 & 0.0307 & 0.0412 & 0.0550 \\
1000 & 0.0186 & 0.0216 & 0.0283 & 0.0358 \\
2000 & 0.0129 & 0.0151 & 0.0197 & 0.0246 \\
3000 & 0.0102 & 0.0118 & 0.0155 & 0.0202 \\
4000 & 0.0087 & 0.0101 & 0.0131 & 0.0172 \\
5000 & 0.0073 & 0.0086 & 0.0113 & 0.0147 \\
10000 & 0.0059 & 0.0069 & 0.0089 & 0.0117 \\
50000 & 0.0025 & 0.0034 & 0.0061 & 0.0077 \\
\noalign{\smallskip}\hline
\end{tabular}
\end{table}

Conover \cite{key-26} presents detailed instructions of how to use
the KS table for obtaining a goodness-of-fit estimate. Next, a very
simple practical example will be shown on how to use the techniques
presented in this paper.

\begin{figure}
\resizebox{0.45\textwidth}{!}{%
  \includegraphics{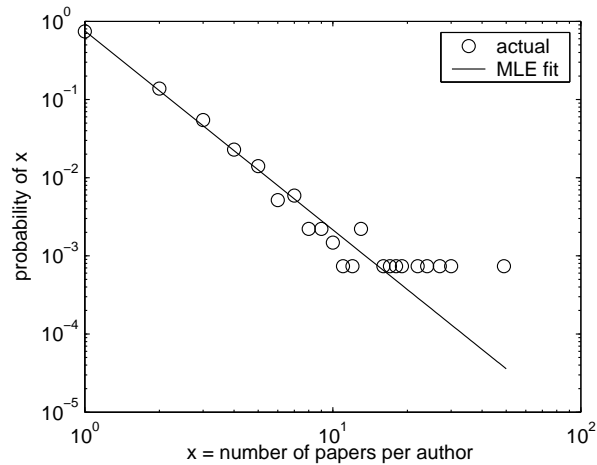}
} \caption{Example of log-log plot of papers per author distribution
for 900 papers in the complex networks field. The circles represent
the empirical distribution and the line represents the MLE estimate
of the power-law distribution. $\gamma_{MLE}=2.544$.}
\label{fig:1}       
\end{figure}

The data set used contains 900 papers in the complex networks field,
and the distribution tested was of the number of papers per author,
often characterized as a power-law known as Lotka's Law
\cite{key-42}. These papers were written and co-written by a total
of 1,354 different authors. Figure \ref{fig:1} shows the empirical
distribution in a log-log plot. The MLE estimation can be obtained
simply by calculating $\frac{1}{N}\sum_{i=1}^{N}\log(x_i)$ given in
Equation \ref{eq:zz}, where $x_{i}$ is the number of papers authored
by author $i$. This sum in this data set equals to 0.2739. By using
Matlab, it is possible to solve Equation \ref{eq:zz} for $\gamma$,
resulting in $\gamma=2.544$. It is also possible to use the table
provided in \cite{key-41}, but it would result in lower precision.
Figure \ref{fig:1} shows also the plot of the fitted power-law line.

Now, in order to test if this fit is reasonable, the KS test can be
used. This test requires the calculation of the maximum distance
between the hypothesized cumulative distribution ($F^{\star}(x)$ - a
power-law distribution with exponent 2.544), and the empirical
distribution $S(x)$. For this case, the test statistic obtained was
$K=0.0117$. The number of samples (number of authors) is $N=1,354$.
The closest value to $N$ in Table \ref{tab:2} is $N=1,000$ (although
it would be statistically "safer" to choose the next highest number
of samples to ensure that the rejection rate is not lower than the
one stated in the test, in practice it is considered reasonable to
approximate to the closest value when the statistic is not close to
the critical values). Looking at the quantile values of the row for
1,000 samples, the observed $K$, 0.0117, is below 0.0186,  the 0.9
quantile. This means that the observed significance level (OSL)  is
greater than 10\%, i.e., in more than 10\% of the cases where the
distribution is an actual power-law, the K statistic is greater than
0.0117.  Therefore, with an OSL greater than 10\%, there is
insufficient evidence to reject the hypothesis that the distribution
is a power-law.

This simple example shows how easy the calculation of the MLE
estimate and the $K$ statistic is, and how to consult the KS table
to obtain good basis to confirm or reject the power-law distribution
hypothesis.

\section{Conclusions}
\label{sec:3} A simple experiment using a random deviate generator
shows that linear-fit based methods for estimating the power-law
exponent tend to produce erroneous results. MLE based estimates,
which are simple to produce using tables or built-in math functions
in computational software, provide a more robust estimation of the
power-law exponent.

In conjunction with the MLE method, the KS-type test table given
here can be used to produce a quantitative assessment of
goodness-of-fit, allowing researchers to meaningfully assess and
compare the appropriateness of modeling empirical data as a
power-law distribution.

\end{document}